\documentclass[aps,10pt,pra,twocolumn,showpacs,floatfix,preprintnumbers]{revtex4-1}
\usepackage{amsmath,amssymb}
\usepackage{graphicx}

\newcommand{\slitsep}{d} 
\newcommand{\slitwidth}{a}
\newcommand{\tot}{\text{cm}} 
\newcommand{\IN}{\mathcal{I}_N}

\begin{document}

\title{Nonlocal Young tests with Einstein-Podolsky-Rosen correlated particle pairs}

\author{Clemens Gneiting}
\affiliation{Albert-Ludwigs University of Freiburg, Hermann-Herder-Stra{\ss}e 3, 79104 Freiburg, Germany}
\author{Klaus Hornberger}
\affiliation{University of Duisburg-Essen, Lotharstra{\ss}e 1-21, 47048 Duisburg, Germany}

\date{\today}

\begin{abstract}
We evaluate the nonlocal spatial interference displayed by Einstein-Podolsky-Rosen entangled particle pairs after they pass through a double-grating arrangement. 
An entanglement criterion is derived which serves to certify the underlying entanglement only from the observed spatial correlations.
We discuss the robustness of the scheme along with a number of possible realizations with matter waves.
\end{abstract}


\preprint{\textsf{published in Phys.~Rev.~A~{88}, 013610 (2013)}}
\maketitle

\section{Introduction}

Interferometry plays a central role in physics, with applications ranging from sensitive phase measurements, e.g.\ to monitor the spatial displacements experienced by gravitational wave detectors \cite{abadie2011gravitational}, to fundamental tests of quantum physics, such as the wave-particle duality of increasingly complex quantum objects \cite{cronin2009optics,hornberger2012colloquium}.
Given the power of interference experiments, it is natural to ask how their scope can be extended to access entanglement---the second pillar of nonclassicality \cite{Einstein1935a}, and an important resource in quantum information \cite{Horodecki2009,tichy2011essential,Weedbrook2012}. Such a combined witnessing of spatial interference and entanglement would not only amount to a striking demonstration of the departure of non-local quantum behavior from classical physics. It might also be used for entanglement-enhanced metrological applications such as phase estimation schemes or quantum lithography \cite{Boto2000a, DAngelo2001a}.

So far all experiments that combined entanglement with spatial interference were based on photons \cite{Boto2000a, DAngelo2001a, Taguchi2008a, Horne1989a, PhysRevLett.64.2495}. This is due to their great practical use in information science, and above all to the existence of a mature technology for producing and manipulating photonic systems. However, the recent advances in the control of ultracold atoms suggest that it will be possible to carry out similar experiments with material particles. In particular, tailored Einstein-Podolsky-Rosen (EPR) entangled atom pairs can be produced by dissociating Feshbach molecules \cite{Kheruntsyan2005a, Kheruntsyan2006a,Gneiting2008a, Gneiting2010b} or by colliding Bose-Einstein condensates 
\cite{Perrin2007a,RuGway2011a,Bucker2011a,Kofler2012a}, in a process similar to parametric down conversion of laser light, the established method to generate entanglement among photons \cite{Reid2009a}.

It would be a great experimental advancement to demonstrate a nonlocal spatial interference effect with particles of matter. This would establish the presence of both entanglement and the wave-particle duality in a single experiment on tangible material objects, allowing one to transfer potential application schemes
such as quantum lithography  \cite{Boto2000a, DAngelo2001a}
from photons to the realm of quantum matter. 
Moreover, interpretational issues such as the nonlocality of Bohmian trajectories could be addressed \cite{PhysRevLett.110.060406}. 
However, it is not obvious whether the observation of an interference pattern in the coincidence signal of two detectors already proves the existence of entanglement. This is the more so, as any experiment will be characterized by a non-ideal EPR source and other imperfections leading to a reduced fringe visibility of the interference signal.

\begin{figure}[b]
\includegraphics[width=\columnwidth]{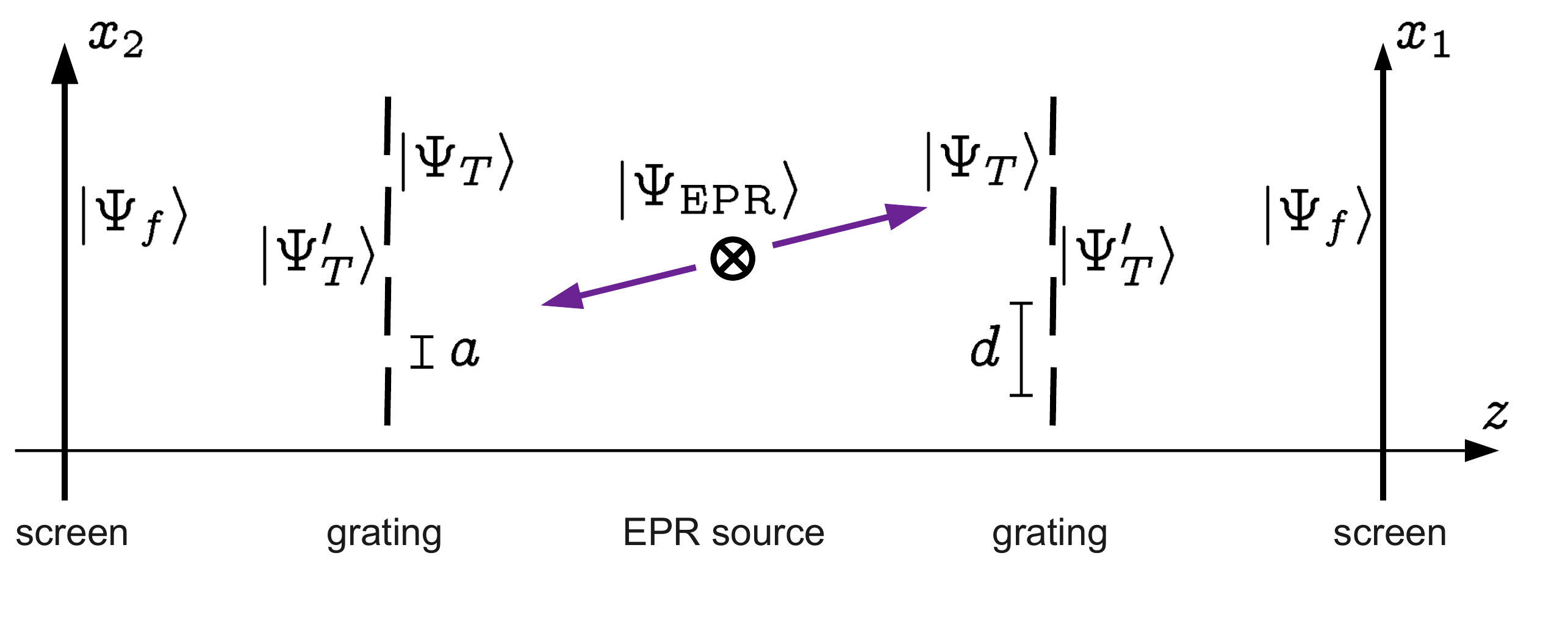}
\caption{\label{SchematicSetup} Generic setup of a nonlocal Young experiment with EPR entangled particle pairs. A source provides EPR entangled particle pairs of equal mass, which are then separately passed through a grating or double slit  (with slit separation $\slitsep$ and slit width $\slitwidth$).  While the position measurements yield an unstructured pattern on each screen, interference fringes are expected if one evaluates the correlated outcomes $x_1+x_2$. Taking the EPR correlation to be in the transversal $x$-direction it is sufficient to describe the longitudinal $z$-motion, which separates the particles, by classical counterpropagating trajectories.}
\end{figure}

In this article we provide an entanglement criterion for nonlocal spatial interference based on EPR entangled particle pairs, and we work out the conditions for the successful detection of a nonlocal interference signal. A schematic  generalizing the single-particle Young experiment is depicted in Fig.~\ref{SchematicSetup}. A source creates a pair of EPR entangled particles which travel freely into opposite directions until each one passes through a grating or double slit. After a further free evolution their positions are recorded by spatially resolving detectors located at opposite sides of the experiment. Even for a source emitting ideal EPR particle pairs no interference will be observed at each of the detectors. However, a `nonlocal' interference pattern is expected to emerge if one analyzes the combined detection records at both sides by focusing on the center of mass of the coincident pairs.

A major challenge in such nonlocal interference experiments lies in verifying the presence of continuous variable entanglement. Once the gratings are passed the two-particle state of motion is strongly non-Gaussian, with a broadened and structured momentum distribution. The standard entanglement criteria \cite{Duan2000a,Simon2000a,Braunstein2005} can then not be used, even though they apply to arbitrary states, since they will detect entanglement only if the corresponding position and momentum variances are sufficiently squeezed. Moreover, only position measurements are easily doable with material particles, practically ruling out tomographic techniques of entanglement verification.
Recently, we described  a viable method based on \emph{modular variables}, which  captures situations where entanglement gives rise to spatial interference \cite{Gneiting2011a}. In the following we introduce the corresponding criterion, adapted to the specific correlations displayed by the EPR interference experiments.

The article is structured as follows: In Sect.~\ref{sec:nli} we describe how the nonlocal interference pattern can be calculated for finite EPR sources (i.e.~sources that produce EPR states with finite variances in all coordinates and momenta), allowing us to discuss the requirements and conditions for observing nonlocal interference. After briefly explaining the concept of modular variables, the entanglement criterion  is then formulated in Sect.~\ref{sec:iev}, along with a discussion of its robustness. In Sect.~\ref{sec:emwi} we discuss different experimental scenarios, before presenting our conclusions in Sect.~\ref{sec:conc}.

\section{Nonlocal spatial interference from Einstein-Podolsky-Rosen correlations}\label{sec:nli}

\subsection{The normalized EPR state}
Einstein, Podolsky, and Rosen considered originally  
the idealized state \cite{Einstein1935a}
\begin{equation}
\int \mathrm{d}x |x\rangle_1 |x\rangle_2 = \int \mathrm{d}p |p\rangle_1 |-p\rangle_2.\nonumber
\end{equation}
Switching to center-of-mass coordinates, and disregarding normalization, it may be written as 
\begin{equation}
\int 
\mathrm{d}x_{\text{rel}} 
\mathrm{d}p_{\text{cm}} 
\,\delta(p_{\text{cm}}) \delta(x_{\text{rel}}) 
|x_{\text{rel}}\rangle_{\text{rel}}
|p_{\text{cm}}\rangle_{\text{cm}} 
.\nonumber
\end{equation}
The state supports perfect correlations both in the relative position $\mathsf{x}_{\text{rel}} = \mathsf{x}_1-\mathsf{x}_2$ and in the center-of-mass momentum $\mathsf{p}_{\text{cm}} = \mathsf{p}_1+\mathsf{p}_2$. 
The expectation values of these commuting observables vanish for the idealized EPR state, as do their 
variances  $\sigma_{x, \text{rel}}^2$ and  $\sigma_{p, \text{cm}}^2$. This implies that the conjugate relative momentum $\mathsf{p}_{\text{rel}} = (\mathsf{p}_1-\mathsf{p}_2)/2$ and the center-of-mass position $\mathsf{x}_{\text{cm}} = (\mathsf{x}_1+\mathsf{x}_2)/2$ remain undetermined.

\begin{figure}
\includegraphics[width=\columnwidth]{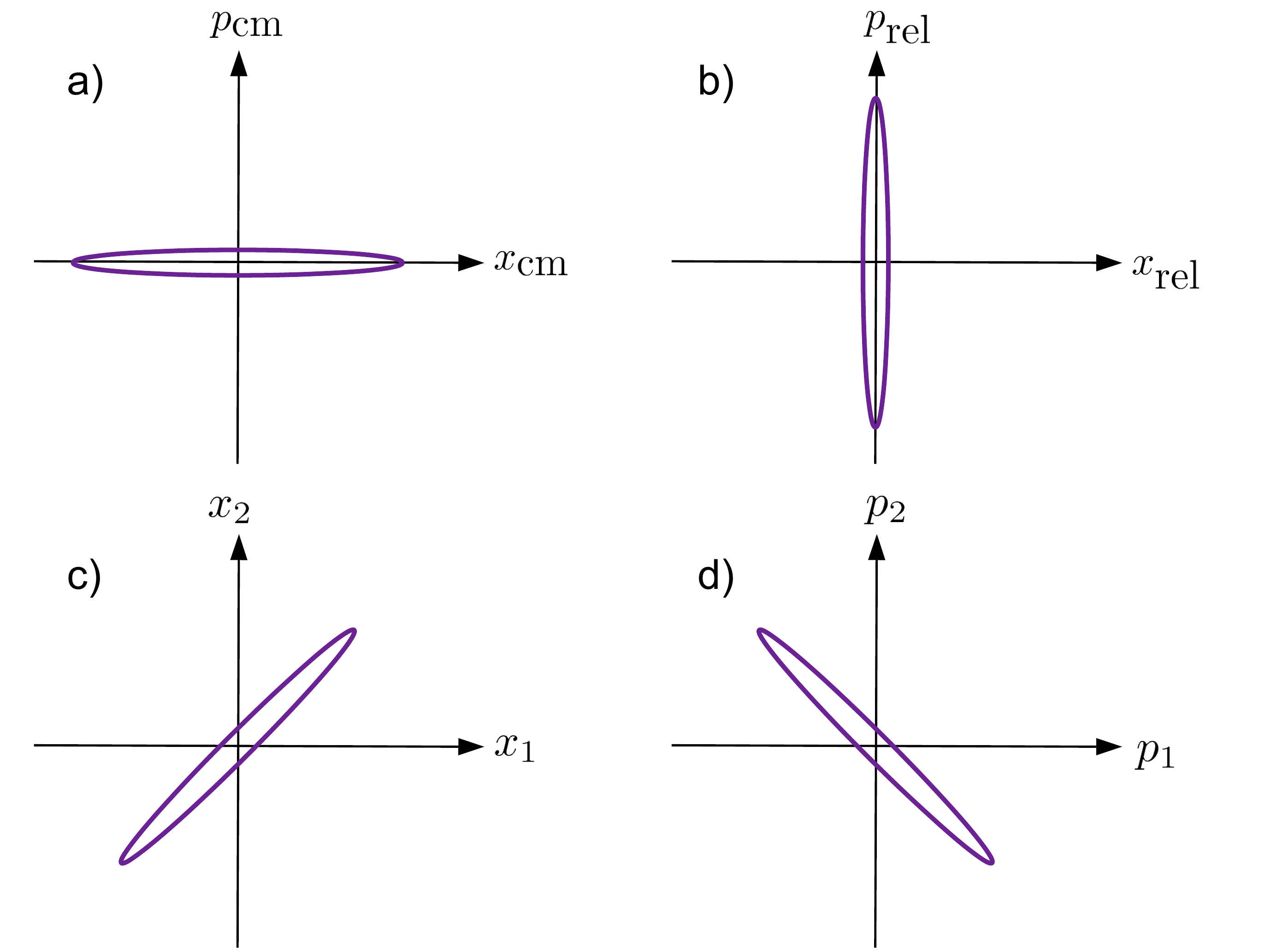}
\caption{\label{EPR_state} Correlations expressed by the squeezed Gaussian (\ref{RealisticEPRstate}) representing a normalizable EPR state. The sketches display two-dimensional projections of the four-dimensional Wigner function, emphasizing characteristic properties of the EPR state. (a) The conjugate relationship between the squeezed center-of-mass momentum $p_{\rm cm}$ and the correspondingly broad center-of-mass coordinate $x_{\rm cm}$. (b) The conjugate relation between the relative coordinate and its momentum. (c, d) Correlations in the particle coordinates and momenta due to the squeezing in $x_{\rm rel}$ and $p_{\rm cm}$.}
\end{figure}

The idealized EPR state  is readily generalized to a normalized squeezed Gaussian wavefunction exhibiting finite position and momentum variances,
\begin{eqnarray} \label{RealisticEPRstate}
|\Psi_{\text{EPR}}\rangle & = & 
\frac{1}{\sqrt{{2 \pi} \sigma_{p, \text{cm}}\sigma_{x, \text{rel}}}}
\int \mathrm{d}x_{\text{rel}} \mathrm{d}p_{\text{cm}} |x_{\text{rel}}\rangle_{\text{rel}} |p_{\text{cm}}\rangle_{\text{cm}} \nonumber \\
 & & \times \mathrm{exp}\left({-\frac{p_{\text{cm}}^2}{4 \sigma_{p, \text{cm}}^2}}{-\frac{x_{\text{rel}}^2}{4 \sigma_{x, \text{rel}}^2}}\right). 
\end{eqnarray}
The resulting correlations are sketched in Figure \ref{EPR_state} for both $\sigma_{x, \text{rel}}$ and  $\sigma_{p, \text{cm}}$ much smaller than the uncertainties of an (unsqueezed) minimum uncertainty state.

\subsection{Interferometric two-particle evolution}

We proceed to determine the joint probability distribution 
for recording the two EPR particles on detection screens after each  passed a grating with  $N$ slits. To this end it is helpful to divide the evolution into three steps: (i) the free time evolution from the source to the gratings, (ii) the instantaneous effect of passing through the gratings, and (iii) a further period of free time evolution from the gratings to the screens. This can be done since all the correlations probed in such a setting reside in the transversal motion of the two particles. In the Fraunhofer approximation the longitudinal motion from the source to the screens may be viewed as taking place with definite velocities which are to be averaged in the end. The longitudinal position can thus be considered as a parametrization of time.

We take the longitudinal direction to be the  $z$-axis and the grating bars to be aligned along the $y$-axis such that the relevant motion takes place in the $x$-direction. The free time evolution of the Gaussian wavefunction (\ref{RealisticEPRstate}) from the source to the gratings can be determined analytically. Denoting the arrival time of the particles at the gratings as $T$, the evolved state immediately before passing the gratings reads as
\begin{eqnarray} \label{StateBeforeGrids}
|\Psi_T\rangle & = & 
\frac{{\rm e}^{{\rm i}\varphi_T}}{\sqrt{\mathcal{N}_T}}
\int \mathrm{d}x_{\text{cm}}\mathrm{d}x_{\text{rel}} \,
|x_{\text{cm}}\rangle_{\text{cm}} |x_{\text{rel}}\rangle_{\text{rel}}
\nonumber \\
 & & \times 
  \mathrm{exp}\left({-\frac{x_{\text{cm}}^2}{4 \sigma_{x, \text{cm}}^2 \xi_{T, \text{cm}}}}{-\frac{x_{\text{rel}}^2}{4 \sigma_{x, \text{rel}}^2 \xi_{T, \text{rel}}}}\right) \, .
\end{eqnarray}
Here we transformed from the center-of-mass momentum $p_{\text{cm}}$ to the center-of-mass position $x_{\text{cm}}$, with $\sigma_{x, \text{cm}} = \hbar/(2 \sigma_{p, \text{cm}})$ the corresponding (large) uncertainty. The evolution introduces additional phase factors and results in a dispersion-induced broadening of the original Gaussian wave packets described by the complex dispersion factors 
\begin{eqnarray}
\xi_{t, \text{cm}} &=& 1+ \frac{\mathrm{i} \hbar t}{4 \sigma_{x, \text{cm}}^2 m},
\\
\xi_{t, \text{rel}} &=& 1+ \frac{\mathrm{i}\hbar t}{ \sigma_{x, \text{rel}}^2 m},
\end{eqnarray}
associated with total mass $ 2m$ and reduced mass $ m/2$, respectively. The normalization factor $\mathcal{N}_T$ and the global phase $\varphi_T$, which are readily calculated, will not be required in the following.

The effect of traversing the gratings can be captured by the grating operators 
\begin{equation} \label{GridOperators}
\mathsf{G}_i = 
\sum_{n \in \IN}
\int_{-\frac{\slitwidth}{2}}^{\frac{\slitwidth}{2}} \mathrm{d}x |n \slitsep+x\rangle_i \langle n \slitsep+x|_i,
\end{equation}
which are projectors acting on the individual particles, $i = 1,2$.
The gratings are taken to consist of $N$ slits, with slit distance $\slitsep$ and slit width $\slitwidth$, 
see Fig.~\ref{SchematicSetup}.
The slit indices are taken from
\begin{equation}
 \mathcal{I}_N=\left\{ -\frac{N-1}{2}, -\frac{N-3}{2},\ldots, \frac{N-1}{2}\right\},
\end{equation}
which guarantees that the gratings are arranged symmetrically with respect to the $z$-axis for all $N$. The gratings are assumed to be ideal in the sense that imperfections and the dispersion force between particle and  grating can be neglected. This is permissible because such slit imperfections would not affect the fringe structure of the nonlocal interference pattern, but only its envelope.

Immediately after traversing the gratings the state follows from the projection $ \mathsf{G}_1 \otimes \mathsf{G}_2 |\Psi_T\rangle$. Switching to the particle coordinates $x_1 = x_{\text{cm}}+x_{\text{rel}}/2$ and $x_2 = x_{\text{cm}}-x_{\text{rel}}/2$ the entangled two-particle state thus takes the form
\begin{widetext}
\begin{eqnarray} \label{StateAfterGrids}
|\Psi_T'\rangle & = & 
\frac{\mathrm{e}^{\mathrm{i}\varphi_T'}}{\sqrt{\mathcal{N}_T'}}
\sum_{n,n' \in \IN} \int_{-\frac{\slitwidth}{2}}^{\frac{\slitwidth}{2}} \mathrm{d}x 
\int_{-\frac{\slitwidth}{2}}^{\frac{\slitwidth}{2}} \mathrm{d}x' 
   \mathrm{exp}\left({-\frac{([n+n'] \slitsep+x+x')^2}{16 \sigma_{x, \text{cm}}^2 |\xi_{T, \text{cm}}|^2}}-\frac{([n-n'] \slitsep+x-x')^2}{4 \sigma_{x, \text{rel}}^2 |\xi_{T, \text{rel}}|^2}\right) \nonumber\\
 & & \times \mathrm{exp}\left({\mathrm{i} \frac{\sigma_{p, \text{cm}}^2 ([n+n'] \slitsep+x+x')^2 T}{16 m \sigma_{x, \text{cm}}^2 \hbar |\xi_{T, \text{cm}}|^2}} +\mathrm{i} \frac{\sigma_{p, \text{rel}}^2 ([n-n'] \slitsep+x-x')^2 T}{m \sigma_{x, \text{rel}}^2 \hbar |\xi_{T, \text{rel}}|^2}\right) \; |n \slitsep+x\rangle_1 |n' \slitsep+x'\rangle_2.
\end{eqnarray}
\end{widetext}

\subsection{Conditions for nonlocal interference}

We can now identify the conditions for observing nonlocal spatial interference at the screens. 
As follows from the general discussion of two-particle correlations in \cite{Gneiting2011a}, nonlocal spatial interference requires that the state prepared by the gratings is correlated with respect to the slits traversed. That is, if we measure which slit a particle took on one side, by detecting it right after the grating, we must be able to infer which slit the other particle took on the other side. This is guaranteed by requiring that
\begin{equation} \label{SlitCorrelationCondition}
\sigma_{x, \text{rel}} |\xi_{T, \text{rel}}| \ll \slitsep,
\end{equation}
because the second Gaussian in Eq.~(\ref{StateAfterGrids}) can then be well approximated by a Kronecker delta for the slit indices $n$ and $n'$ (along with with a factor describing irrelevant intra-slit correlations). 

The slit correlation condition (\ref{SlitCorrelationCondition}) combines two requirements. On the one hand, the initial state must be sufficiently squeezed in the relative position, $\sigma_{x, \text{rel}} \ll \slitsep$. On the other hand, to guarantee (\ref{SlitCorrelationCondition}) at the relevant time $T$ when the gratings are passed, the  dispersive broadening of the wave packet evolving from the source to the screen must remain sufficiently bounded.
The propagation time $T$ from the source to the gratings thus cannot exceed
\begin{equation} \label{PropagationTime}
T_{\rm max} = \sigma_{x, \text{rel}}^2 m/\hbar \ll  \slitsep^2 m/\hbar.
\end{equation}
This shows that there is a tradeoff between the squeezing of the relative position $\sigma_{x, \text{rel}}$ and the maximum admissible propagation time from the source to the screen. This is no problem of principle since the longitudinal motion might be uncoupled from the transverse dynamics, allowing one to propagate the particles with arbitrary velocity to their gratings. In practice, one may well be forced to start out with an isotropic two-particle state, using apertures for defining the longitudinal direction. In this case the dispersion tradeoff  (\ref{PropagationTime})
can be a quite severe restriction.

As a second condition one must require that the $N$ slits are  illuminated uniformly by the incoming wavefunction. This is ensured by the first Gaussian in Eq.~(\ref{StateAfterGrids}) provided
\begin{equation} \label{EqualIlluminationCondition}
\sigma_{x, \text{cm}} |\xi_{T, \text{cm}}| \gtrsim N \slitsep.
\end{equation}
Since the slit correlation condition (\ref{SlitCorrelationCondition}) already requires that the relative dispersion remains modest, $T \lesssim  T_{\rm max}$, we can estimate 
\begin{equation}
| \xi_{T, \text{cm}} |\leq
\sqrt{1+\left( \frac{\sigma_{x, \text{rel}}}{2\sigma_{x, \text{cm}}}\right)^4} \simeq 1,
\end{equation}
because $\sigma_{x, \text{rel}} \ll \sigma_{x, \text{cm}}$. The initial state (\ref{RealisticEPRstate}) must therefore exhibit a large  center-of-mass uncertainty $\sigma_{x, \text{cm}} \gtrsim N \slitsep$.

\subsection{The modular momentum entangled state}

Applying the conditions (\ref{SlitCorrelationCondition}) and (\ref{EqualIlluminationCondition}) to Eq.~(\ref{StateAfterGrids}) yields a greatly simplified expression for the slit-entangled two-particle state present once both gratings have been passed,
\begin{eqnarray} \label{StateAfterGridsApproximated}
|\Psi_T'\rangle & \approx & 
\frac{\mathrm{e}^{\mathrm{i}\varphi_T'}}{\sqrt{\mathcal{N}_T'}}
\int_{-\frac{\slitwidth}{2}}^{\frac{\slitwidth}{2}} \mathrm{d}x \int_{-\frac{\slitwidth}{2}}^{\frac{\slitwidth}{2}} \mathrm{d}x' 
\mathrm{exp}\left({-\frac{(x-x')^2}{4 \sigma_{x, \text{rel}}^2 \xi_{T, \text{rel}}}}\right)
\nonumber \\
 & & \times  \sum_{n \in\IN} |n \slitsep+x\rangle_1 |n \slitsep+x'\rangle_2.
\end{eqnarray}
Specifically, condition (\ref{SlitCorrelationCondition}) implies that those contributions to the wave function where the two particles do not pass opposite slits with $n'=n$ can be neglected. For instance, the next neighbor contributions with $n-n'=\pm 1$ are weighted by the factor $\exp[-(d/(2 \sigma_{x, {\rm rel}} |\xi_{T, \text{rel}}|))^2]$, and already a moderate ratio of $d/(\sigma_{x, {\rm rel}} |\xi_{T, \text{rel}}|)=5$ suppresses these by three orders of magnitude compared to the opposite-slit contributions. Condition (\ref{EqualIlluminationCondition}), on the other hand, effects that all opposite slit pairs  contribute equally to the resulting superposition. Since the contributions at the margins of the gratings are diminished by the factor $\exp[-(N d/(4 \sigma_{x, {\rm cm}} |\xi_{T, \text{cm}}|))^2]$, already a ratio of $N d/(\sigma_{x, {\rm cm}} |\xi_{T, \text{cm}}|)=1$ limits the amplitude decrease toward the margins to 7\%.

The state (\ref{StateAfterGridsApproximated}) describes a superposition of the particle pair passing through opposite slits. This becomes most transparent once we rewrite the state in position representation,
\begin{equation} \label{ModularMomentumEntangledState}
\langle x_1, x_2|\Psi_T'\rangle = 
\frac{1}{\sqrt{N}}
\sum_{n \in\IN} \langle x_1-n \slitsep, x_2-n 
\slitsep|\Psi_s\rangle.
\end{equation}
Here the normalized wavefunction $|\Psi_s\rangle$ describes that both particles are confined to a single pair of opposite slits,
\begin{eqnarray} \label{OppositeSlitPairState}
\langle x_1, x_2|\Psi_s\rangle & = &
\frac{\mathrm{e}^{\mathrm{i}\varphi_s}}{\sqrt{\mathcal{N}_s}} \, \mathrm{exp}\left({-\frac{(x_1-x_2)^2}{4 \sigma_{x, \text{rel}}^2 \xi_{T, \text{rel}}}}\right) \chi_a(x_1)\chi_a(x_2),
\nonumber \\ &&
\end{eqnarray}
with $\chi_a(x)=\Theta\left(\frac{\slitwidth}{2}+x\right)\Theta\left(\frac{\slitwidth}{2}-x\right)$. 
As before, the phase $\varphi_s$ and the normalization factor $\mathcal{N}_s$ will not be required in the following.

The wavefunction (\ref{ModularMomentumEntangledState}) is an instance of a \emph{modular momentum entangled} (MME) state, a class of states discussed in a more general context in \cite{Gneiting2011a}. As for any MME state, the momentum representation of (\ref{ModularMomentumEntangledState}) reads
\begin{equation} \label{MMEMomentumRepresentation}
\langle p_1, p_2|\Psi_T'\rangle = \frac{1}{\sqrt{N}} \sum_{n \in\IN} \mathrm{e}^{-\mathrm{i} (p_1+p_2) n \slitsep/\hbar} \langle p_1, p_2|\Psi_s\rangle.
\end{equation}
This in turn implies a nonlocal interference behavior if the particle momenta are measured.
The joint momentum probability distribution takes the form
\begin{equation} \label{MomentumProbabilityDistribution}
|\langle p_1, p_2|\Psi_T'\rangle|^2 = |\langle p_1, p_2|\Psi_s\rangle|^2 F_N\left(\frac{(p_1+p_2) \slitsep}{h}\right),
\end{equation}
where the interference pattern is captured by the fringe function
\begin{equation} \label{FringeFunction}
F_N(\xi) = 1+\frac{2}{N} \sum_{j=1}^{N-1} (N-j) \cos(2 \pi j \xi).
\end{equation}
It reduces to a sinusoidal fringe pattern in the case of double slits, while $F_N(x)$ develops sharpened main maxima and suppressed side maxima for $N>2$. Note that the period of the fringe pattern is given by the ``grating momentum'' $h/\slitsep$.

A distinct interference pattern can only emerge if the envelope in (\ref{MomentumProbabilityDistribution}), as given by the momentum distribution $|\langle p_1, p_2|\Psi_s\rangle|^2$, varies slowly over the extension of a single period $h/\slitsep$, and if it is sufficiently broad to cover several fringes. These conditions are met in the present case since the width of the momentum distribution of (\ref{OppositeSlitPairState}) is essentially determined by single-slit diffraction, i.~e.~by $h/\slitwidth$. This is always greater than the grating momentum $h/\slitsep$, since $\slitsep > \slitwidth$. More precisely, the envelope is determined by a convolution of the opposite slit pair contributions and the correlated relative motion,
\begin{eqnarray}
\langle p_1, p_2|\Psi_s\rangle & = & \frac{\mathrm{e}^{\mathrm{i} \tilde\varphi_s}}{\sqrt{\tilde{\mathcal{N}}_s}} \int \mathrm{d}\tilde{p}\,\mathrm{sinc}\left({ \frac{[p_1+p_2+\tilde{p}] \slitwidth}{2 \hbar}}\right)
 \nonumber\\
 & & \times   
\mathrm{sinc}\left({\frac{[p_1+p_2-\tilde{p}] \slitwidth}{2 \hbar}}\right) 
  \\
& & \times\mathrm{exp}\left[{-\xi_{T, \text{rel}} \left(\frac{[p_1-p_2-\tilde{p}] \sigma_{x, \text{rel}}}{2 \hbar}\right)^2}\right]. \nonumber
\end{eqnarray}

\subsection{Far-field interference pattern}

The discussed momentum interference effect is easily observed by letting the particles propagate freely for a time $T_2$ from the gratings to remote detection screens on each side. Denoting the final state of the particles as $|\Psi_f\rangle$,
the joint spatial detection probability 
is directly determined by
the momentum distribution (\ref{MomentumProbabilityDistribution}) of 
$|\Psi'_T \rangle$,
\begin{equation}\label{farfield}
|\langle x_1, x_2|\Psi_f \rangle|^2 = \frac{m^2}{T_2^2} \left|\Big\langle \frac{m x_1}{T_2}, \frac{m x_2}{T_2} \Big| {\Psi}'_T\Big\rangle
\right|^2.
\end{equation}
This assumes that the screens are placed sufficiently far away from the gratings such that one is in the dispersion-dominated limit, $T_2 \gg m N^2 \slitsep^2/\hbar$. 

The position measurements 
at the detection screens may thus be viewed as effective momentum measurements on $|\Psi'_T \rangle$. It follows that the joint spatial probability distribution reproduces the nonlocal momentum interference pattern (\ref{MomentumProbabilityDistribution}),
\begin{eqnarray} \label{FinalStateSpatialProbabilityDistribution}
|\langle x_1, x_2| \Psi_f \rangle|^2 & = & \frac{m^2}{T_2^2} \left|
\Big\langle \frac{m x_1}{T_2}, \frac{m x_2}{T_2} \Big| {\Psi}_s\Big\rangle
\right|^2 \nonumber \\
 & & \times F_N \left( { \frac{m (x_1+x_2) }{T_2 h/\slitsep}}\right).
\end{eqnarray}
This result exhibits the expected nonlocal interference behavior. No fringe pattern will be visible if one looks at either of the screens since the integration over the unobserved particle position will 
remove the fringe function in (\ref{MomentumProbabilityDistribution}),
leaving only the broad envelope determined by $| {\Psi}_s\rangle$.  
Only by recording the coincident detections at both screens and by collecting
the center-of-mass positions
$x_1+x_2$ will an interference pattern emerge.

This proves that it is possible to establish nonlocal interference by exposing EPR entangled particle pairs to gratings, and in this sense to perform an entangled Young experiment. The expected spacing of the fringe pattern is given by $ T_2 h/(m \slitsep)$.

We identified the slit correlation condition (\ref{SlitCorrelationCondition}) and the uniform slit illumination (\ref{EqualIlluminationCondition}) as requirements for a successful implementation. 
In the following section, we will show how the nonlocal spatial interference pattern (\ref{FinalStateSpatialProbabilityDistribution}) can serve as the basis for a rigorous verification of the underlying entanglement.

\section{Interferometric entanglement verification}\label{sec:iev}

As impressive as the correlations expressed by the nonlocal interference pattern (\ref{FinalStateSpatialProbabilityDistribution}) may be, it is not clear \emph{a priori} that they cannot just as well emerge from a classically correlated quantum state, without resorting to entanglement. To exclude this possibility, one must testify the presence of entanglement with a suitable entanglement criterion. Ideally, this should not require measurements beyond the ordinary position measurements giving rise to the interferometric correlations of the EPR Young experiment; in particular we should avoid the necessity of an unfeasible continuous variable state tomography.

\subsection{Modular variables.}

Such an entanglement verification can be achieved using an entanglement criterion in terms of modular variables \cite{Gneiting2011a}. The latter prove useful to capture spatial interference phenomena \cite{Aharonov1969a,Tollaksen2010a,Popescu2010a,Plastino2010a}. They formally decompose the position and momentum operators into step-like integer components $\mathsf{N}_x$ and $\mathsf{N}_p$ and sawtooth-like modular components $\mathsf{\overline{x}}$ and $\mathsf{\overline{p}}$,
\begin{equation}
\mathsf{x} = \mathsf{N}_x \slitsep + \mathsf{\overline{x}}, \hspace{5mm} \mathsf{p} = \mathsf{N}_p \frac{h}{\slitsep} + \mathsf{\overline{p}}.
\end{equation}
Expressed in the position eigenbasis the modular and the integer position are thus given by
\begin{eqnarray}
\mathsf{\overline{x}} &=& \int_{-\infty}^{\infty} \mbox{d} x\, \overline{x}(x) |x\rangle \langle x|,
\\
\mathsf{N}_{x} &=& \int_{-\infty}^{\infty} \mbox{d} x\,  \frac{x-\overline{x}(x)}{\slitsep} |x\rangle \langle x|,
\end{eqnarray}
with
\begin{eqnarray}
\overline{x}(x) &:=& \left(x+\frac{\slitsep}{2}\right) \mbox{mod}\, \slitsep - \frac{\slitsep}{2}.
\end{eqnarray}
The momentum operators are defined similarly  by the spectral function
\begin{eqnarray}\label{pop}
\overline{p}(p) &:=& \left(p+\frac{h}{2 \slitsep}\right) \mbox{mod}\, \frac{h}{\slitsep} - \frac{h}{2 \slitsep}.
\end{eqnarray}
Note that the standard position and momentum eigenvectors $|x\rangle$ and $|p\rangle$ can also be interpreted as the joint eigenstates of the respective integer and modular observable. The latter can thus be deduced from ordinary position and momentum measurements.

\subsection{Moments and variances}

Interpreted in terms of the modular variables, the correlations displayed by the MME state (\ref{ModularMomentumEntangledState}) describe reduced variances for the total modular momentum $\mathsf{\overline{p}}_{\tot} = \mathsf{\overline{p}}_1+\mathsf{\overline{p}}_2$. This is the reason for naming it {\it modular momentum entangled}.  Also the spread of the relative integer position $\mathsf{N}_{x, \text{rel}} = \mathsf{N}_{x, 1}-\mathsf{N}_{x, 2}$ is reduced. In particular, if the slit correlation condition (\ref{SlitCorrelationCondition}) is satisfied the variance of $\mathsf{N}_{x, \text{rel}}$ vanishes by construction, as a result of the opposite slit pair correlations,
\begin{eqnarray} \label{MMeVarianceRelativeIntegerPosition}
\langle (\Delta \mathsf{N}_{x, \text{rel}})^2 \rangle & = & \langle \Psi_T'| \mathsf{N}_{x, \text{rel}}^2 |\Psi_T' \rangle - (\langle \Psi_T'| \mathsf{N}_{x, \text{rel}} |\Psi_T' \rangle)^2 \nonumber \\
 & = & 0.
\end{eqnarray}
Using (\ref{ModularMomentumEntangledState}) and (\ref{OppositeSlitPairState}) and noting $N_x(x) = (x-\overline{x}(x))/\slitsep$ one finds that in fact all moments $m \geq 1$ vanish,
\begin{eqnarray}
\langle \mathsf{N}_{x, \text{rel}}^m \rangle & = & \int_{-\infty}^{\infty} \mbox{d}x_1 \int_{-\infty}^{\infty} \mbox{d}x_2 [N_x(x_1)-N_x(x_2)]^m \nonumber \\
 & & \times |\langle x_1, x_2|\Psi_T' \rangle|^2 \nonumber \\
 & = & \frac{1}{N} \sum_{n_1,n_2\in\IN} \int_{n_1 \slitsep-\frac{\slitwidth}{2}}^{n_1 \slitsep+\frac{\slitwidth}{2}} \mbox{d}x_1 \int_{n_2 \slitsep-\frac{\slitwidth}{2}}^{n_2 \slitsep+\frac{\slitwidth}{2}} \mbox{d}x_2 \nonumber \,\delta_{n_1 n_2} \\
 & & \times (n_1-n_2)^m |\langle x_1-n_1 \slitsep, x_2-n_2 \slitsep|\Psi_s \rangle|^2 \nonumber \\
 & = & 0.
\end{eqnarray}
On the other hand, in the relevant limit $\slitwidth \ll \slitsep$ and using (\ref{MomentumProbabilityDistribution}), the moments of the total modular momentum are given by
\begin{eqnarray} \label{TotalModularMomentumMoments}
\langle \mathsf{\overline{p}}_{\tot}^m \rangle & = & \langle \Psi_T'| (\mathsf{\overline{p}}_1 + \mathsf{\overline{p}}_2)^m |\Psi_T' \rangle \nonumber \\
 & = & \int_{-\infty}^{\infty} \mbox{d} p_1 \int_{-\infty}^{\infty} \mbox{d} p_2 \, [\overline{p}(p_1)+\overline{p}(p_2)]^m \nonumber \\
  & & \times |\langle p_1, p_2|\Psi_s\rangle|^2 F_N\left({ \frac{(p_1+p_2) \slitsep}{h}}\right) \nonumber \\
 & = & \frac{\slitsep^2}{h^2}  \int_{-h/2 \slitsep}^{h/2 \slitsep} \mbox{d} p_1 \int_{-h/2 \slitsep}^{h/2 \slitsep} \mbox{d} p_2 \, (p_1+p_2)^m \nonumber \\
 & & \times F_N\left({ \frac{(p_1+p_2) \slitsep}{h}}\right).
\end{eqnarray}
Here we used the scale separation between the width of the fringe pattern envelope and its period; this permits one to apply the approximation 
\begin{equation}
 \int_{-\infty}^{\infty} \mbox{d}x \, \chi(x) E(x) \approx \int_{-\infty}^{\infty} \mbox{d}x' E(x')  \int_{-\lambda/2}^{\lambda/2} \frac{\mbox{d}x}{\lambda} \, \chi(x), 
\end{equation}
which is valid for a $\lambda$-periodic function $\chi(x)$ and an envelope function $E(x)$ that varies slowly over the extent of a single period $\lambda$.

Putting the fringe function (\ref{FringeFunction}) into (\ref{TotalModularMomentumMoments}), one obtains for the first two moments 
\begin{eqnarray}
 \langle \mathsf{\overline{p}}_{\tot} \rangle &=& 0,
 \\
 \label{p2cm}
 \langle \mathsf{\overline{p}}_{\tot}^2 \rangle &=& \frac{h^2}{6 \slitsep^2} (1-S_2(N)).
\end{eqnarray}
The  positive function $S_2$, also found in \cite{Gneiting2011a}, is defined by
\begin{equation} \label{SqueezingFunction}
S_2(N) = \frac{6}{\pi^2} \sum_{j=1}^{N-1} \frac{N-j}{N j^2}.
\end{equation}
It is bounded by $1>S_2(N)$, increases monotonically, and is well approximated by its asymptotic form
\begin{equation} \label{SqueezingFunctionasymptotic}
S_2(N) \sim 1-\frac{6 (1+\gamma+\ln(N))}{\pi ^2 N},
\end{equation}
involving Euler's constant $\gamma\simeq 0.577$. 
The variance of the total modular momentum 
\begin{equation} \label{MMeVarianceTotalModularMomentum}
\langle (\Delta \mathsf{\overline{p}}_{\tot})^2 \rangle = \frac{h^2}{6 \slitsep^2} (1-S_2(N))
\end{equation}
thus decreases with a growing number $N$ of slits.
For large $N$ this squeezing of the total modular momentum scales as
\begin{equation}
 \langle (\Delta \mathsf{\overline{p}}_{\tot})^2 \rangle \sim \left(\frac{h}{\pi d}\right)^2
 \frac{1+\gamma+\ln(N)}{N}.
\end{equation}

\subsection{Post-measurement analysis}

In the proposed Young test the experimenter performs ordinary position measurements directly behind the gratings and in the far field, yielding the joint probability densities ${\rm prob}(x_1,x_2)$ and  ${\rm prob}(p_1,p_2)$, respectively.
The required modular variances 
$\langle (\Delta \mathsf{N}_{x, \text{rel}})^2 \rangle$ 
and 
$\langle (\Delta \mathsf{\overline{p}}_{\tot})^2 \rangle$ 
are then obtained by a post-measurement analysis of this data. 

Specifically, the moments of the relative integer position are obtained by evaluating
\begin{eqnarray}
 \langle\mathsf{N}^m_{x,\rm rel} \rangle
 &=& \int_{-\infty}^{\infty} \mbox{d}x_1
 \int_{-\infty}^{\infty} \mbox{d}x_2
 [N_x(x_1)-N_x(x_2)]^m \nonumber\\
 &&\times{\rm prob}(x_1,x_2),
\end{eqnarray}
with $N_x(x)=(x-\overline{x}(x))/d$.
Similarly, one calculates
\begin{eqnarray}
 \langle\mathsf{p}^m_{\rm cm} \rangle
 &=& \int_{-\infty}^{\infty} \mbox{d}p_1
 \int_{-\infty}^{\infty} \mbox{d}p_2
 [\overline{p}(p_1)+\overline{p}(p_2)]^m \nonumber\\
 &&\times{\rm prob}(p_1,p_2),
\end{eqnarray}
where $\overline{p}(p)$ is defined in (\ref{pop}). 
From an operational point of view, this is all that is required to test the entanglement criterion  (\ref{ModularEntanglementCriterion}).

\subsection{Shifted modular variables}

Carrying out the post-measurement analysis one can always choose the  position and momentum coordinates in such a way that the maxima of the interference pattern coincide with vanishing values of the corresponding modular variable. This reflects the optimal choice and was the case in our calculations so far.

The general case can be modeled by introducing an additional phase  $\varphi$ into (\ref{FinalStateSpatialProbabilityDistribution}) which shifts the interference pattern. The expression (\ref{TotalModularMomentumMoments}) for the moments of the total modular momentum then becomes
\begin{eqnarray}
\langle \mathsf{\overline{p}}_{\tot}^m \rangle & = & \frac{\slitsep^2}{h^2}  \int_{-h/2 \slitsep}^{h/2 \slitsep} \mbox{d} p_1 \int_{-h/2 \slitsep}^{h/2 \slitsep} \mbox{d} p_2 \, (p_1+p_2)^m \nonumber \\
 & & \times F_N\left({ \frac{(p_1+p_2) \slitsep}{h}}+\varphi \right).
\end{eqnarray}
This results in the modified variance 
\begin{equation}
\langle (\Delta \mathsf{\overline{p}}_{\tot})^2 \rangle = \frac{h^2}{6 \slitsep^2} (1-S_{2}(N,\varphi)),  
\end{equation}
where 
\begin{equation}
S_{2}(N,\varphi) = \frac{6}{\pi^2} \sum_{j=1}^{N-1} \frac{N-j}{N j^2} \cos(j \varphi) < S_{2}(N).        
\end{equation}
A finite $\varphi$ can result in a substantial deterioration of the total modular momentum squeezing, while the moments of the relative integer positions remain unaffected. In the remainder we consider again the case $\varphi = 0$. This is no restriction of generality due to  the freedom of choice of $\varphi$ in the post-measurement analysis.

\subsection{Modular entanglement criterion}

The reduced fluctuations in the integer relative position and the total modular momentum can be used to verify unambiguously the underlying entanglement. This is achieved with an entanglement criterion similar to the modular entanglement criterion derived in \cite{Gneiting2011a}, where the squeezing was considered to occur in a different set of two-particle observables, namely the modular relative position $\overline{x}_{\text{rel}} = \overline{x}_1-\overline{x}_2$ and the total integer momentum $N_{p, \tot} = N_{p,1}+N_{p,2}$. 

In the present case the relevant entanglement criterion  reads
\begin{equation} \label{ModularEntanglementCriterion}
\frac{\slitsep^2}{h^2} \langle (\Delta \mathsf{\overline{p}}_{\tot})^2 \rangle_{\rho} + \langle (\Delta \mathsf{N}_{x, \text{rel}})^2 \rangle_{\rho} < 2 \mathcal{C}_{\bar{\mathsf{p}},\mathsf{N}_x}.
\end{equation}
Any state satisfying this condition must  be entangled. Note that the criterion (\ref{ModularEntanglementCriterion}) is sufficient but not necessary. The constant $\mathcal{C}_{\bar{\mathsf{p}},\mathsf{N}_x}$ is given by the smallest root $\mu_0$ of the equation
\begin{equation}
 \frac{\rm d}{{\rm d}x}
 \left[
 {\rm e}^{- \pi x^2}
 M\left(- \frac{\pi}{2}\mu+\frac{1}{4}, \frac{1}{2}, 2 \pi x^2\right)\right]_{x = 1/2}=0,
\end{equation}
with
$M(a, b; x)$ the Kummer function. Numerical evaluation yields $\mathcal{C}_{\bar{\mathsf{p}},\mathsf{N}_x} \cong 0.078 \, 235$.

The MME state (\ref{ModularMomentumEntangledState}) satisfies the entanglement criterion (\ref{ModularEntanglementCriterion}) for any $N \ge 2$, as follows directly from
the variances (\ref{MMeVarianceRelativeIntegerPosition}) and (\ref{MMeVarianceTotalModularMomentum}),
\begin{equation}
\frac{1}{6} (1-S_2(N))+0 < 2 \mathcal{C}_{\bar{\mathsf{p}},\mathsf{N}_x}.
\end{equation}
This proves that it is indeed possible to detect unambiguously  entanglement based on the nonlocal interference which is produced by exposing EPR entangled particle pairs to a Young-like grating setup. Already for the least sensitive entanglement scheme, the case $N=2$ of a double slit on each side, the squeezing function (\ref{SqueezingFunction}) evaluates as $S_2(2) = 0.30$, resulting in a sum of uncertainties staying 25\% below the threshold.

\subsection{Robustness of the entanglement detection}

\subsubsection{ Admixture of a classically slit-correlated state}

To get a generic understanding of the robustness of the entanglement detection scheme with respect to visibility reduction, one may ask how many classical (i.e.~no interference supporting) correlations can be admixed to the EPR state without compromising the criterion (\ref{ModularEntanglementCriterion}). To this end we introduce the  \emph{classically} slit-correlated state
\begin{eqnarray} \label{ClassicallySlitCorrelatedState}
\langle x_1,x_2|\rho_{\text{cl}}|x_1',x_2'\rangle & = & \frac{1}{N} \sum_{n\in\IN} \langle x_1-n \slitsep, x_2-n \slitsep|\Psi_s\rangle \nonumber \\
 & & \times \langle \Psi_s|x_1'-n \slitsep, x_2'-n \slitsep\rangle.
\end{eqnarray}
Compared to the MME state $\rho_{\text{MME}} = |\Psi_T'\rangle \langle \Psi_T'|$ determined by (\ref{ModularMomentumEntangledState}), the state (\ref{ClassicallySlitCorrelatedState}) lacks the coherences between different opposite-slit pairs. It does therefore not exhibit nonlocal interference. It carries the variances $\langle (\Delta \mathsf{N}_{x, \text{rel}})^2 \rangle_{\text{cl}} = 0$ and $\langle (\Delta \mathsf{\overline{p}}_{\tot})^2 \rangle_{\text{cl}} = h^2/(6 \slitsep^2)$; this latter variance of the total modular momentum is the maximum possible, reflecting complete ignorance. 

Let us now consider the  mixture 
\begin{equation} \label{ClassicallySlitCorrelatedMixture}
\rho_w = (1-w) \rho_{\text{MME}} + w \, \rho_{\text{cl}}
\end{equation}
of the MME state (\ref{ModularMomentumEntangledState})
and the classically slit-correlated state (\ref{ClassicallySlitCorrelatedState}) 
with $w\in(0,1)$.
We find that the variances evaluate as $\langle (\Delta \mathsf{N}_{x, \text{rel}})^2 \rangle_w 
= 0$ and
\begin{eqnarray} \label{WernerVarianceTotalModularMomentum}
\langle (\Delta \mathsf{\overline{p}}_{\tot})^2 \rangle_{w} & = & (1-w) \langle (\Delta \mathsf{\overline{p}}_{\tot})^2 \rangle_{\text{MME}} + w \langle (\Delta \mathsf{\overline{p}}_{\tot})^2 \rangle_{\text{cl}} \nonumber \\
 & = & \frac{h^2}{6 \slitsep^2} [1-(1-w) S_2(N)],
\end{eqnarray}
noting that all involved first moments vanish, $\langle \mathsf{N}_{x, \text{rel}} \rangle_{\text{MME}} = \langle \mathsf{N}_{x, \text{rel}} \rangle_{\text{cl}} = \langle \mathsf{\overline{p}}_{\tot} \rangle_{\text{MME}} = \langle \mathsf{\overline{p}}_{\tot} \rangle_{\text{cl}} = 0$.

Comparing (\ref{MMeVarianceTotalModularMomentum}) and (\ref{WernerVarianceTotalModularMomentum}) one sees that the squeezing of the total modular momentum is diminished by the amount of classical admixture $w$. 
This corresponds to a reduced visibility, and the fringe pattern (\ref{MomentumProbabilityDistribution}) gets replaced by 
\begin{equation} 
 |\langle p_1, p_2|\Psi_s\rangle|^2
 \left[w+(1-w)\,F_N\left(\frac{(p_1+p_2) \slitsep}{h}\right)
 \right]   \,.
\end{equation}

In the case of double slits, $N =2$, we thus find that the entanglement criterion (\ref{ModularEntanglementCriterion}) remains satisfied as long as $w < (12 \mathcal{C}_{\bar{\mathsf{p}},\mathsf{N}_x}-1)/S_2(2)+1 = 0.79$. 

In other words, we can admix up to $79 \%$ of a classically correlated state and still detect the entanglement in the blurred fringe pattern. This robustness manifests the power of the modular entanglement detection scheme and it provides a comfortable cushion to deal with potential noise sources and experimental limitations, such as decoherence and a finite detection resolution, which reduce the fringe visibility.

\subsubsection{Admixture of a separable state}

In the opposite case, where the source produces uncorrelated particle pairs, the wavefunction behind the gratings is described by the separable state 
\begin{equation}                                                            
|\Psi_{\rm sep}\rangle=|\psi_{\rm ms}\rangle_1|\psi_{\rm ms}\rangle_2,
\end{equation}
with single-particle multislit states
\begin{equation}
 \langle x|\psi_{\rm ms}\rangle=
 \frac{1}{\sqrt N}\sum_{n\in\IN}\langle x-nd|\psi_{\rm s}\rangle\,.
\end{equation}
Here the state $|\psi_{\rm s}\rangle$ corresponds to the single-particle state prepared by a single slit. The state $|\Psi_{\rm sep}\rangle$ then leads to local interference patterns on each side.  This implies correlations in the modular total momentum which reduce its variance. However, the lacking slit correlations result in a substantial variance of the relative integer position
\begin{equation}
 \langle (\Delta \mathsf{N}_{x,\rm rel})^2\rangle_{\rm sep}
 = \frac{1}{6}(N^2-1)\,,
\end{equation}
in contrast to the vanishing variance (\ref{MMeVarianceRelativeIntegerPosition}).
As one expects, already for $N=2$ this exceeds substantially the threshold value $2\mathcal{C}_{\bar{\mathsf{p}},\mathsf{N}_x}$ of the entanglement criterion (\ref{ModularEntanglementCriterion}).  In other words, a mixed state with a separable admixture exceeding $w=4 \mathcal{C}_{\bar{\mathsf{p}},\mathsf{N}_x}$ (see (\ref{ClassicallySlitCorrelatedMixture})), i.e. of about 31\%, is no longer detected by the entanglement criterion (\ref{ModularEntanglementCriterion}).

\subsubsection{Extended EPR sources}

Another possible reason for a reduced interference visibility are imprecise EPR sources. We therefore discuss in the following how the conditions and results derived above are affected if the initial state is not a pure EPR state (\ref{RealisticEPRstate}) but a mixture of EPR states with mutually displaced centers in phase space. They will be characterized by the phase space coordinates $\Gamma\equiv\big(x_{\text{cm}}^{(0)}, x_{\text{rel}}^{(0)}, p_{\text{cm}}^{(0)}, p_{\text{rel}}^{(0)}\big)$ indicating where each EPR state is initially located with respect to the center-of-mass and relative coordinates:
\begin{eqnarray} \label{Displaced_EPR_state}
|\Psi_{\text{EPR}}^{(\Gamma)}\rangle & = & 
\frac{1}{\sqrt{{2 \pi} \sigma_{x, \text{cm}} \sigma_{x, \text{rel}}}}
\int\mathrm{d}x_{\text{cm}}\mathrm{d}x_{\text{rel}} 
|x_{\text{cm}}\rangle_{\text{cm}} |x_{\text{rel}}\rangle_{\text{rel}}\nonumber \\
 & & \times \mathrm{exp}\left({-\frac{(x_{\text{cm}}-x_{\text{cm}}^{(0)})^2}{4 \sigma_{x, \text{cm}}^2}}{-\frac{(x_{\text{rel}}-x_{\text{rel}}^{(0)})^2}{4 \sigma_{x, \text{rel}}^2}}\right) \nonumber \\
 & & \times \exp\left(\mathrm{i} \frac{p_{\mathrm{cm}}^{(0)} \, x_{\mathrm{cm}}}{\hbar}+\mathrm{i} \frac{p_{\mathrm{rel}}^{(0)} \, x_{\mathrm{rel}}}{\hbar} \right).
\end{eqnarray}
Comparison with Eq.~(\ref{RealisticEPRstate}) shows that the previously considered EPR state is centered at $\Gamma=0$. The general mixture is given by
\begin{align} \label{Imperfect_EPR_state}
\rho_{\mu} = \int \mathrm{d}\Gamma \, \mu(\Gamma) |\Psi_{\text{EPR}}^{(\Gamma)}\rangle \langle \Psi_{\text{EPR}}^{(\Gamma)}|,
\end{align}
with 
$\mathrm{d}\Gamma=\mathrm{d}x_{\text{cm}}^{(0)}\mathrm{d} x_{\text{rel}}^{(0)}\mathrm{d} p_{\text{cm}}^{(0)}\mathrm{d} p_{\text{rel}}^{(0)}$. It is thus determined by the probability distribution function $\mu(\Gamma)$, taken in the following to be a Gaussian centered at the origin, which is fully characterized by the standard deviations ${\sigma}^{(0)}_{x, \rm cm}$, 
${\sigma}^{(0)}_{p, \rm cm}$, 
${\sigma}^{(0)}_{x, \rm rel}$, and 
${\sigma}^{(0)}_{p, \rm rel}$.

To see the effect of the mixing (\ref{Imperfect_EPR_state}), we first determine the 
interference pattern of a (moderately) displaced, pure EPR state (\ref{Displaced_EPR_state}). Freely propagating the wavefunction  $|\Psi_{\text{EPR}}^{(\Gamma)}\rangle$
 for time $T$ and then through the gratings yields
\begin{widetext}
\begin{eqnarray} \label{Displaced_EPR_state_after_grids}
|\Psi_T'^{(\Gamma)}\rangle & = & 
\frac{\mathrm{e}^{\mathrm{i}\varphi_T'}}{\sqrt{\mathcal{N}_T'}}
\sum_{n,n' \in\IN} \int_{-\frac{\slitwidth}{2}}^{\frac{\slitwidth}{2}} \!\!\!\mathrm{d}x
\int_{-\frac{\slitwidth}{2}}^{\frac{\slitwidth}{2}}\!\!\! \mathrm{d}x' \,  |n \slitsep+x\rangle_1 |n' \slitsep+x'\rangle_2\,
 \mathrm{exp}\left(\mathrm{i} \, \phi_{\text{cm}}\left(\frac{n+n'}{2} \slitsep+\frac{x+x'}{2}\right) + \mathrm{i} \, \phi_{\text{rel}}([n-n'] \slitsep+x-x')\right)
 \nonumber\\
 & & \times 
 \mathrm{exp}\left({-\frac{([n+n'] \slitsep+x+x'-2x_{\text{cm}}^{(0)}-{p_{\text{cm}}^{(0)}} T/m)^2}{16 \sigma_{x, \text{cm}}^2 |\xi_{T, \text{cm}}|^2}}-\frac{([n-n'] \slitsep+x-x'- x_{\text{rel}}^{(0)}-2{p_{\text{rel}}^{(0)}}T/{m})^2}{4 \sigma_{x, \text{rel}}^2 |\xi_{T, \text{rel}}|^2}\right)
 \; .
\end{eqnarray}
Here, we introduced the phase functions for the center-of-mass and relative motion:
\begin{align}
\phi_{\text{cm}}(x) &= \frac{p_{\text{cm}}^{(0)} \, x}{\hbar |\xi_{T, \text{cm}}|^2} + \frac{\hbar T (x-x_{\text{cm}}^{(0)})^2}{16m \sigma_{x, \text{cm}}^4 |\xi_{T, \text{cm}}|^2},  \\
\phi_{\text{rel}}(x) &= \frac{p_{\text{rel}}^{(0)} \, x}{\hbar |\xi_{T, \text{rel}}|^2} + \frac{\hbar T (x-x_{\text{rel}}^{(0)})^2}{4m \sigma_{x, \text{rel}}^4  |\xi_{T, \text{rel}}|^2}. 
\end{align}
\end{widetext}
Comparing the form (\ref{Displaced_EPR_state_after_grids}) of the displaced wavefunction with (\ref{StateAfterGrids})  it follows that the slit correlation condition (\ref{SlitCorrelationCondition}) and the requirement of uniform illumination (\ref{EqualIlluminationCondition}) remain necessary conditions.

In the following, we determine what additional constraints must be satisfied by the displacements $\Gamma$  for a successful Young test. First, it must be guaranteed that the grating is still illuminated uniformly. Noting that the wavefunction (\ref{Displaced_EPR_state_after_grids}) 
is centered around the classical displacements 
\begin{align}
 x_{\text{cm}}^{(T)}&=x_{\text{cm}}^{(0)}+\frac{p_{\text{cm}}^{(0)}T}{2m},
\\
x_{\text{rel}}^{(T)}&=x_{\text{rel}}^{(0)}+2\frac{p_{\text{rel}}^{(0)}T}{m},
\end{align}
we get the requirement 
\begin{equation} \label{r1}
\left| x_{\text{cm}}^{(T)} \right|\ll N \slitsep \,.
\end{equation}

Next, consider the impact of the displacements on the resulting slit correlations. To this end, it is helpful to express $x_{\text{rel}}^{(T)}$ in modular variables, $x_{\text{rel}}^{(T)}=N_{x,\text{rel}}^{(T)} \slitsep+\overline{x}_{\text{rel}}^{(T)}$.
Similarly to the undisplaced case, the first Gaussian in (\ref{Displaced_EPR_state_after_grids}), combined with the slit correlation condition (\ref{SlitCorrelationCondition}), implies ideal correlations, $n-n'=N_{x,\text{rel}}^{(T)}$.
Note that in contrast to the case $\Gamma=0$, we must now also take into account that it is not necessarily opposite slit pairs that are correlated.
Uniform illumination of the slits on both sides demands that the offset $N_{x,\text{rel}}^{(T)} \slitsep$ is small compared to the extension of the grating $Nd$, \begin{equation}\label{r2}
N_{x,\text{rel}}^{(T)} \ll N.                                                                                                                                                                   
\end{equation}
Moreover, to make sure that both particles can pass the gratings in spite of their correlation the modular part must satisfy
\begin{equation}\label{r3}
\left|\overline{x}_{\text{rel}}^{(T)} \right|< \frac{\slitwidth}{2}.
\end{equation}

If the conditions (\ref{r1})--(\ref{r3}) are met, we obtain the interference pattern
\begin{align} \label{Displaced_EPR_interference}
|\langle p_1,p_2|\Psi_T'^{(\Gamma)}\rangle|^2 =& |\langle p_1,p_2|\Psi_s\rangle|^2 \\
 & \times F_{N'}\left(\frac{[p_1+p_2] \slitsep}{h}-\frac{p_{\text{cm}}^{(0)} \, \slitsep}{h |\xi_{T,\text{cm}}|^2}\right), \nonumber
\end{align}
with an interference order $N'=N-|N_{x,\text{rel}}^{(T)}|$. 
It implies that the interference is remarkably robust against phase-space displacements, since only a shift in the center-of-mass momentum, $p_{\text{cm}}^{(0)}$, directly affects the phase of the nonlocal interference pattern. 

With this we are in a position to discuss the implications for the mixed state (\ref{Imperfect_EPR_state}). The center-of-mass requirement (\ref{r1}) leads to the constraints
${\sigma}^{(0)}_{x, \rm cm}\ll Nd$ and 
${\sigma}^{(0)}_{p, \rm cm}\ll mNd/T$, and the condition (\ref{r2}) for the 
relative motion implies
${\sigma}^{(0)}_{x, \rm rel}\ll Nd$ and 
${\sigma}^{(0)}_{p, \rm rel}\ll mNd/T$.
While these ``classical'' requirements are relatively easy to meet, the sensitivity of the interference pattern 
(\ref{Displaced_EPR_interference}) with respect to phase averaging demands a 
significantly tightened control of ${\sigma}^{(0)}_{p, \rm cm}$. 

Specifically, the blurred interference pattern due to the phase averaging results in an increased variance of the total modular momentum as compared to (\ref{p2cm}),
\begin{eqnarray}\label{p2cmred}
 \langle \mathsf{\overline{p}}_{\tot}^2 \rangle_{{\mu}} &=& \frac{h^2}{6 \slitsep^2} \left[1-
 \exp\left(-\frac{\big({\sigma}^{(0)}_{p, \rm cm}d\big)^2}{2 h^2|\xi_{T,\text{cm}}|^2}\right)
 S_2(N)\right].\phantom{aa}
\end{eqnarray}
A significant reduction of the fringe visibility is thus to be expected once the total momentum spread ${\sigma}^{(0)}_{p, \rm cm}$ exceeds the grating momentum $h/d$. 
This indicates the level of control of the initial state required for a successful entanglement detection.

\subsubsection{Suboptimal EPR states}

Finally, we discuss to what extent one can relax the slit correlation condition (\ref{SlitCorrelationCondition}) and the condition for uniform illumination (\ref{EqualIlluminationCondition}) and still fulfill the modular entanglement criterion (\ref{ModularEntanglementCriterion}), i.e.~we consider suboptimal EPR states with variances that do not sufficiently satisfy (\ref{SlitCorrelationCondition}) and (\ref{EqualIlluminationCondition}). In that case, the state prepared by the gratings cannot be approximated by the MME state (\ref{ModularMomentumEntangledState}), but must instead be replaced by
\begin{align} \label{Suboptimal_EPR_state}
\langle x_1,& x_2|\Psi_T'\rangle \nonumber \\
= & \frac{1}{\sqrt{\mathcal{N}}} \sum_{n, n' \in \IN}
\mathrm{exp}\left({-\frac{([n+n'] \slitsep)^2}{16 \sigma_{x, \text{cm}}^2 |\xi_{T, \text{cm}}|^2}}
-\frac{([n-n'] \slitsep)^2}{4 \sigma_{x, \text{rel}}^2 |\xi_{T, \text{rel}}|^2}\right) \nonumber \\
& \times \langle x_1-n \slitsep, x_2-n' \slitsep|\Psi_s\rangle.
\end{align}
Here we still assume that the time of flight $T$ from the source to the gratings is sufficiently small (see (\ref{PropagationTime})) such that the
modifications of the phase described by the second line of Eq.~(\ref{StateAfterGrids}) can be neglected.
Note that in the limiting cases (\ref{SlitCorrelationCondition}) and (\ref{EqualIlluminationCondition}) the Gaussian in (\ref{Suboptimal_EPR_state}) reduces to the Kronecker delta $\delta_{n,n'}$ yielding the MMS state (\ref{ModularMomentumEntangledState}).

Based on the suboptimal EPR state (\ref{Suboptimal_EPR_state}), we can investigate the sum of variances in the modular entanglement criterion (\ref{ModularEntanglementCriterion}) as a function of the widths $\sigma_{x, \text{cm}}$ and $\sigma_{x, \text{rel}}$. Again, one finds that the entanglement detection is remarkably robust against suboptimal realizations of the uncertainties of the EPR state. Given $\sigma_{x, \text{cm}} = 1.5 N d$ (i.e.~(\ref{EqualIlluminationCondition}) is satisified), the critical values $\sigma_{x, \text{rel}}^{(\text{crit})}$  where the left-hand side of (\ref{ModularEntanglementCriterion}) reaches the entanglement detection threshold are shown in Table \ref{Critical_Uncertainties}; they hardly depend on the number of slits $N$ in the investigated range, where the width can increase up to $\sigma_{x, \text{rel}}^{(\text{crit})} \approx 0.46 \, d$ before the entanglement detection fails.

Similarly, for $\sigma_{x, \text{rel}} = 0.1 d$ (i.e.~(\ref{SlitCorrelationCondition}) is satisfied), the critical center-of-mass uncertainties $\sigma_{x, \text{cm}}^{(\text{crit})}$ where the interference order is reduced from $N$-slit interference to $(N-1)$-slit interference are located  approximately at $\sigma_{x, \text{cm}}^{(\text{crit})} \approx 0.15 N d$ for all considered $N$, cf.~Table \ref{Critical_Uncertainties}. The case of two slits, $N=2$, is excluded here since it exhibits full two-slit interference for any $\sigma_{x, \text{cm}}$.

\begin{table}[tb] 
  \begin{tabular}{l|lllllll}
    $N$ & 2 & 3 & 4 & 5 & 10 & 20 & 30\\ \hline
    $\sigma_{x, \text{rel}}^{(\text{crit})}/d$ & 0.462 & 0.459 & 0.458 & 0.457 & 0.457 & 0.457 & 0.457\\
    $\sigma_{x, \text{cm}}^{(\text{crit})}/(N d)$ & \phantom{0}--\phantom{0} & 0.148 & 0.146 & 0.148 & 0.140 & 0.127 & 0.119
  \end{tabular}
  \caption{\label{Critical_Uncertainties}Critical uncertainties $\sigma_{x, \text{rel}}^{(\text{crit})}$ and $\sigma_{x, \text{cm}}^{(\text{crit})}$ of the EPR state required for a successful entanglement detection in the $N$-slit experiment, given in units of the slit separation $d$. $\sigma_{x, \text{rel}}^{(\text{crit})}$ denotes the maximal uncertainty in the relative coordinate in compliance with a successful entanglement detection (with fixed $\sigma_{x, \text{cm}} = 1.5 N d$). $\sigma_{x, \text{cm}}^{(\text{crit})}$ denotes the center-of-mass uncertainty where the interference order is reduced from $N$-slit interference to $(N-1)$-slit interference (with fixed $\sigma_{x, \text{rel}} = 0.1 d$).} 
\end{table}

\section{Experimental implementations}
\label{sec:emwi}

Let us now turn to possible experimental demonstrations of nonlocal Young tests and of the verification of entanglement based on EPR correlated particle pairs. We will see that conceivable implementations are quite diverse, which is a result of the generic nature of the presented scheme.

\subsection{Photon experiments}

While this article focuses on entangled matter waves, it should be emphasized that nonlocal interference can be observed with photons as well. Here one can rely on established methods for generating EPR entangled photon pairs, e.g. by parametric down conversion of a laser beam \cite{PhysRevLett.68.3663,PhysRevLett.92.210403}. Such bipartite interference experiments have been performed in different contexts, for instance in quantum lithography \cite{DAngelo2001a}, ghost interferometry \cite{DAngelo2004a}, or the spatial implementation of qubits \cite{Taguchi2008a}. These experiments demonstrated nonlocal spatial interference, but they 
could not verify the continuous variable entanglement for want of a rigorous criterion. 

The analysis in Sects.~\ref{sec:nli} and \ref{sec:iev} can be carried over to EPR entangled photon pairs because the interferometric arrangement allows one to treat the photons as distinguishable particles, to be recorded at distinct positions $x_1$, $x_2$ on a photodetector.
Moreover, the Kirchhoff diffraction integral in Fraunhofer approximation for the photonic modes yields the same expressions as the free propagator of a quantum particle in paraxial approximation. All that needs to be done is to express the evolved time $t$ in terms of the longitudinal momentum $p_z=zm/t$, which is given for the photons by $p_z=h/\lambda$.

In a recent article  \cite{Carvalho2012a} Carvalho et al. describe an experiment with down-converted entangled photons. They report  a significant 
observation of entanglement with double slits, based on the criterion (\ref{ModularEntanglementCriterion}) obtained from \cite{Gneiting2011a}.

\begin{figure}[tb]
\includegraphics[width=\columnwidth]{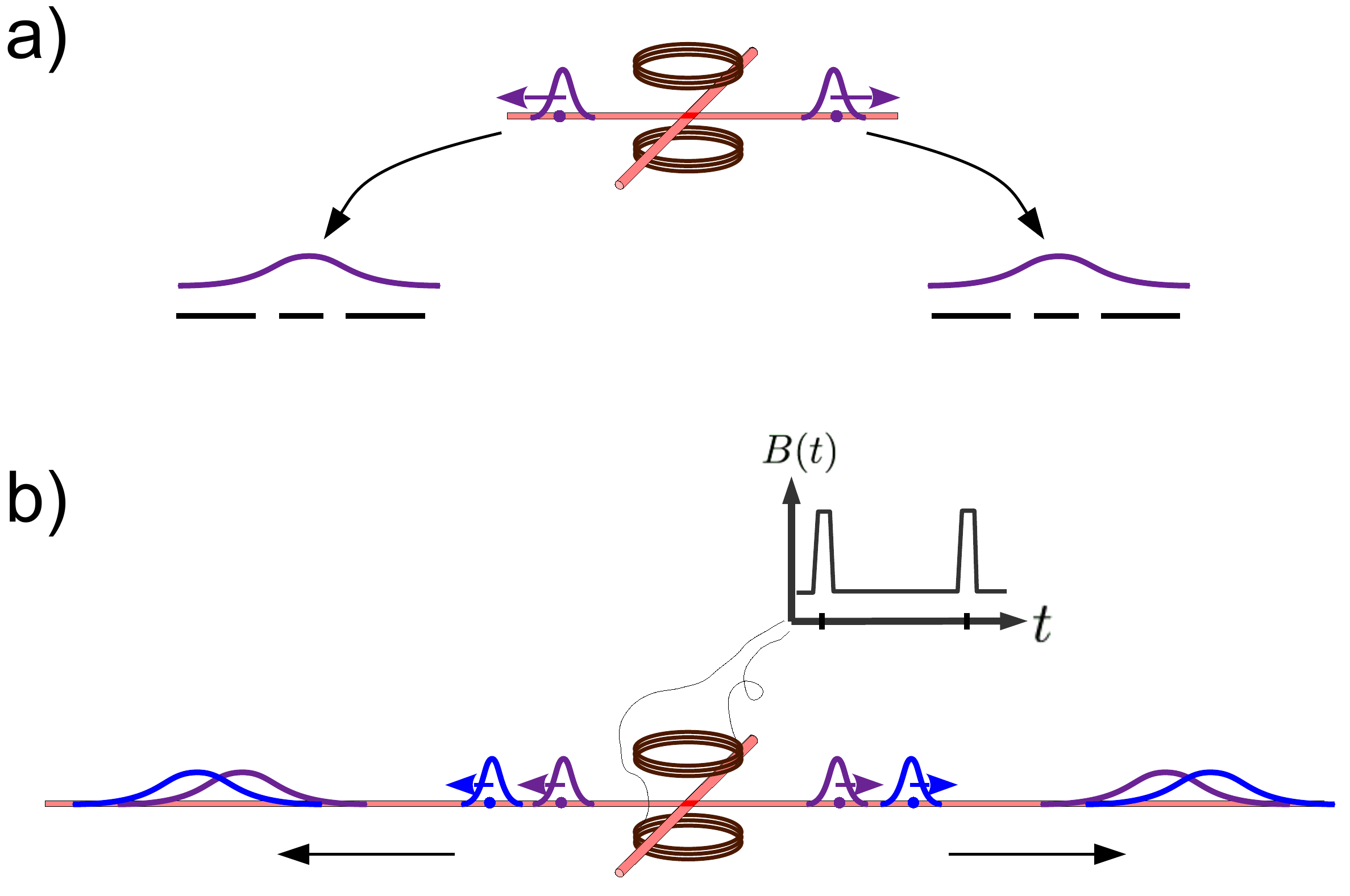}
\caption{\label{FeshbachDissociation} Schematic setups for realizations of an entangled Young experiment based on the controlled dissociation of ultracold Feshbach molecules. An ultracold Feshbach molecule \cite{RevModPhys.78.1311} initially trapped by two crossing laser beams is dissociated by a magnetic-field pulse, which produces an EPR entangled particle pair. (a) The laser guide is switched off after the completion of the dissociation so that the atoms fall freely toward the gratings on each side. (b) A sequence of two magnetic-field dissociation pulses generating a dissociation-time entangled particle pair \cite{Gneiting2008a,Gneiting2010b}. Once the early and the late wave packets overlap due to the dispersive time evolution, a nonlocal interference pattern can emerge in the recorded positions of the two particles.}
\end{figure}

\subsection{EPR pairs from atomic BECs}

A recent proposal by Kofler et al.\ \cite{Kofler2012a} sets out to produce  EPR entangled atom pairs of metastable helium atoms. This is based on a four-wave mixing process in a Bose-Einstein condensate as in \cite{Perrin2007a}. The helium atoms are kicked against each other by stimulated Raman transitions and collide by $s$-wave scattering. If the applied laser pulses are sufficiently weak, such that on average only a single pair of atoms is detected in the end, the resulting (radial) two-particle state is well described by Eq.~(\ref{RealisticEPRstate}). The pair then falls freely under gravity, each of the atoms traversing a double slit aperture, until they hit the detector where they are recorded with high efficiency and resolution.

As in most of the  entanglement tests based on the criterion  (\ref{ModularEntanglementCriterion}), the detector should be movable, to be placed alternately either directly behind the slits or sufficiently far away. 
When positioned close to the slits it serves to ascertain that the particle pair is sufficiently correlated with respect to the slits traversed. This is quantified by calculating the variance of  $\mathsf{N}_{x, \text{rel}} = \mathsf{N}_{x, 1}-\mathsf{N}_{x, 2}$ from the observed positions.
When positioned in the far field, the detector records essentially the transverse momenta $\mathsf{{p}}_1$, $\mathsf{{p}}_2$ of the particles. 
After correlating the data of both particles
the resulting nonlocal interference pattern (\ref{farfield}) 
will have a finite contrast which limits how well-defined the phase of the  pattern is. This phase uncertainty is quantified by the variance of 
the modular  part of the total momentum $\mathsf{\overline{p}}_{\tot} = \mathsf{\overline{p}}_1+\mathsf{\overline{p}}_2$, the second ingredient to the entanglement criterion.

\subsection{EPR pairs from molecular Feshbach dissociation} 

Another possibility to generate clouds of EPR entangled atom pairs is to use the controlled dissociation of Bose-Einstein condensed Feshbach molecules \cite{RevModPhys.78.1311,Kheruntsyan2005a, Kheruntsyan2006a, Gneiting2008a,Gneiting2010b}. Starting from a sufficiently dilute condensate and applying weak dissociation pulses, it is again possible to focus on single EPR atom pairs by removing multiple pair events in a post-selection procedure. 
A detailed investigation of this dissociation scheme, including the confining geometry induced by trap and guiding lasers, can be found in \cite{Gneiting2010b,Gneiting2010a}.
Using the well-developed techniques of atom interferometry \cite{cronin2009optics} one would then have to implement the required gratings by material or light-induced structures.
A schematic of a possible setup is given in Fig.~\ref{FeshbachDissociation} (a).
The horizontal propagation is induced by the dissociation process, while gravitation causes a vertical acceleration towards the gratings. The relevant transverse EPR correlations thus reside in the horizontal motion.
The horizontal detection screens (not shown) would then have to be vertically movable from close to the gratings to the far field, as described above.

\subsection{Dissociation-time entanglement}

One can also conceive schemes that go without gratings, by directly producing the modular momentum entangled  states. 
Here we discuss a method based on dissociation-time entanglement  \cite{Gneiting2008a,Gneiting2010b}, the  sequential dissociation of a Feshbach molecule at two different times. A sequence of two dissociation pulses can  generate a dissociation state where the atom pair is described by a coherent superposition of an early and a late wave-packet component associated with the two dissociation pulses. The two counter-propagating atoms are thus correlated in the dissociation times, see Fig.~\ref{FeshbachDissociation} (b). 

The early and late wave packet components are spatially separated but propagate with equal velocities.
They are thus described by a modular momentum entangled wavefunction  with $N=2$, similar to (\ref{ModularMomentumEntangledState}).
The MME state is here realized in the longitudinal motion which separates the particles. The two dissociation-time components thus take the role of the slit components prepared by double slits, while the dispersive time evolution leads eventually to the overlap of the two wave packets. 

Position measurements in the overlap regions can  be implemented by resonant photon scattering. The joint probability for the particle position then exhibits a nonlocal interference pattern. This completes the analogy with the Young double slit experiment. To prove entanglement one needs again a complementary measurement; in this case one must detect the positions of the atom pair at a time when the wave packets do not yet overlap to ensure that the particle pair is correlated in the early or late dissociation time.

\subsection{MME states by photon scattering}

In a recent article \cite{Knott2013a} an experiment was proposed which generates essentially a modular momentum entangled matter wave based on photon scattering. A trapped pair of distinguishable, non-interacting, massive particles is illuminated with a plane wave of light. By detecting all scattered and non-scattered photons one gains knowledge about the relative coordinate of the two particles, but not about the center of mass, such that an MME state (\ref{ModularMomentumEntangledState}) with $N=2$ is eventually prepared after about 150 photon detections.

If the two particles are then released from the trap, such that they evolve freely and drop towards a detection screen, one expects a nonlocal spatial interference pattern similar to (\ref{FinalStateSpatialProbabilityDistribution}). An appropriate modular entanglement criterion can then serve to deduce the underlying entanglement from the measured correlations, if one complements the detection by correlation measurements taken briefly after releasing the particles from the trap. However, unlike in the previous proposal, both particles are detected on the same screen, and therefore no macroscopic spatial separation is achieved between the two particles.

\section{Conclusions}\label{sec:conc}

We discussed a generic scheme to generalize the Young interference experiment for the case of two entangled particles, where nonlocal spatial interference is achieved by subjecting each particle to a grating structure. The corresponding quantum state exhibits strongly non-Gaussian continuous variable entanglement, which can be revealed by a variance-based entanglement criterion. The latter is formulated in terms of modular variables, i.e.~coordinates adapted to spatial interference phenomena.

Experimentally, the entanglement detection is based on simple position measurements directly behind the gratings or in the far field; the modular variances are then calculated in a post-measurement analysis.
We find that the entanglement detection scheme is quite robust against noisy EPR sources,   coping even with substantial admixtures of classical correlations and incoherences.
Moreover, while already double slit arrangements allow one to verify entanglement, one can improve the correlations by increasing the number of slits.

We showed that a nonlocal Young test could be performed in a wide range of physical systems.  Its demonstration with material particles would be a striking achievement, demonstrating both the wave-particle duality and the non-locality of quantum mechanics at the same time. 

{\it Acknowledgment:} We thank Maximilian Ebner for helpful comments on the manuscript.


\end{document}